\documentclass[aps,prl,showpacs,twocolumn,amsmath,amssymb]{revtex4-1}
\usepackage{graphicx,bbm}

\newcommand{\be}{\begin{equation}}
\newcommand{\ee}{\end{equation}}

\newcommand{\bra}[1]{{\langle #1 \vert}}

\newcommand{\ket}[1]{{\vert #1 \rangle}}

\newcommand{\ave}[1]{{\langle #1\rangle}}
\newcommand{\ii}{ {\rm i} }
\newcommand{\dd}{ {\rm d} }
\newcommand{\ZZ}{\mathbb{Z}}
\newcommand{\RaR}{\mathbb{R}}
\newcommand{\CC}{\mathbb{C}}

\newcommand{\mm}[1]{{\mathbf{#1}}}

\def\tr{{\,{\rm tr}\,}}
\def\ad{{{\rm ad}\,}}
\def\one{\mathbbm{1}}

\def\im{{\,{\rm Im}\,}}
\newcommand{\La}{{\mathtt L}}
\newcommand{\Ra}{{\mathtt R}}

\begin{document}

\title{Open XXZ spin chain: \\
Nonequilibrium steady state and a strict bound on ballistic transport}
\author{Toma\v{z} Prosen}
\affiliation{Department of Physics, FMF,  University of Ljubljana, Jadranska 19, 1000 Ljubljana, Slovenia
}

\date{\today}

\begin{abstract}
An explicit matrix product ansatz is presented, in the first two orders in the (weak) coupling parameter, for the nonequilibrium steady state of the homogeneous, nearest neighbor Heisenberg XXZ spin 1/2 chain driven by Lindblad operators which act  only at the edges of the chain.
The first order of the density operator becomes in the thermodynamic limit an exact pseudolocal conservation law and yields -- via the Mazur inequality -- a rigorous lower bound on the high-temperature
spin Drude weight. Such a Mazur bound is found a nonvanishing fractal function of the anisotropy parameter $\Delta$ for $|\Delta|<1$.

\end{abstract}

 
\maketitle

{\em Introduction.} 
Exactly solvable models which exhibit certain generic physical properties are of paramount importance in theoretical physics, in particular in condensed matter and statistical physics where one of the key open issues is the transport in low dimensional strongly interacting quantum systems. 
An example par excellence of such models is an anisotropic Heisenberg XXZ spin 1/2 chain with a constant nearest neighbor spin interaction which, in spite of it being Bethe ansatz solvable \cite{giamarchi}, still offers many puzzles. For example, at high temperature and vanishing external magnetic field, it is not clear even if the model exhibits ballistic or diffusive spin transport \cite{affleck}. The question is of long-lasting experimental interest \cite{experiments}. Recently, theoretical study of interacting many-body systems has got a new impetus by invoking the methods of open quantum systems and master equations \cite{breuer} in the study of  quantum transport far from equilibrium \cite{saito,wichterich}. 

We consider the markovian master equation in the Lindblad form \cite{lindblad}
\be
\frac{\dd \rho(t)}{\dd t} = -\ii [H,\rho(t)] + \sum_k 2 L_k \rho(t) L^\dagger_k - \{ L^\dagger_k L_k,\rho(t)\}
\label{eq:lindblad}
\ee
for an open XXZ spin 1/2 chain with the Hamiltonian
\be
H = \sum_{j=1}^{n-1} (2\sigma^+_j \sigma^-_{j+1} + 2\sigma^-_j \sigma^+_{j+1} + \Delta \sigma^{\rm z}_j \sigma^{\rm z}_{j+1})
\ee
where $\sigma^\pm_j = \frac{1}{2}(\sigma^{\rm x}_j\pm\ii\sigma^{\rm y}_j),\sigma^{\rm z}_j,j=1,\ldots,n$ are Pauli operators on a tensor product space
$(\CC^2)^{\otimes n}$, with symmetric Lindblad driving acting on the edges of the chain only
\be
L_{1,2} = \sqrt{(1\pm\mu)\varepsilon/2}\, \sigma^\pm_1,\quad
L_{3,4} = \sqrt{(1\mp\mu)\varepsilon/2}\, \sigma^\pm_n.
\ee
In this Letter we construct an exact {\em nonequilibrium steady state} (NESS) solution of this model in the regime of weak coupling (small $\varepsilon$) with the method which seems to be un-related to the Bethe ansatz and expresses NESS in the form of a homogeneous matrix product operator. Such an ansatz has been employed earlier for exact solutions of classical many-body stochastic processes \cite{blythe}, however with an important distinction that here the constructing auxiliary matrices satisfy {\em cubic} instead of quadratic algebraic relations.  Our solution gives birth to a spin current related conservation law of an infinite chain, which is almost local (pseudolocal) in the metallic regime $|\Delta|< 1$, and hence is used in a Mazur inequality \cite{mazur} to bound the spin Drude weight and prove ballistic transport for certain values of anisotropy $\Delta$, while for other we use efficient numerical computation. We observe that the graph of the Mazur bound versus $\Delta$ exhibits a  fractal structure.

{\em Boundary driven XXZ chain.} NESS $\rho_\infty = \lim_{t\to\infty}\rho(t)$ is a fixed point of the flow (\ref{eq:lindblad})
\be
-\ii (\ad H) \rho_\infty + \varepsilon \hat{\cal D} \rho_\infty = 0,
\label{eq:fixedpoint}
\ee
where $(\ad H)\rho := [H,\rho]$ and
\begin{eqnarray}
\hat{\cal D} &:=& {\textstyle\frac{1}{2}}(1+\mu) \hat{\cal D}_+ +  {\textstyle\frac{1}{2}}(1-\mu)\hat{\cal D}_-, \quad {\rm with} \label{eq:defD}\\
\hat{\cal D}_\pm\rho&:=&
2 \sigma^\pm_1 \rho \sigma^\mp_1 - \{\sigma^\mp_1 \sigma^\pm_1,\rho\}+
2 \sigma^\mp_n \rho \sigma^\pm_n - \{\sigma^\pm_n \sigma^\mp_n,\rho\}. \nonumber
\end{eqnarray}
Note the distinct roles of two bath parameters: {\em coupling strength} $\varepsilon$ controls the strength of coupling to the spin baths at the edges of the chain, 
while {\em driving strength} $\mu$ controls the nonequilibrium forcing due to unequal average spin polarizations (or chemical potentials, in the spinless fermion formulation \cite{giuliano}) 
of the two baths.
For example, in the common derivation of the master eq. (\ref{eq:lindblad}) \cite{breuer} weak coupling (small $\varepsilon$) is a standard assumption, whereas 
the linear response physics would be mimicked by taking small $\mu$ at nonsmall $\varepsilon$ (see e.g. \cite{pz,marko}).

Here we address far-from-equilibrium physics within the regime of weak coupling, so we formally expand NESS in terms of the coupling parameter $\varepsilon$
\be 
\rho_\infty = \sum_{p=0}^\infty (\ii \varepsilon)^p \rho^{(p)}.
\label{eq:pertansatz}
\ee
Plugging the ansatz (\ref{eq:pertansatz}) into the fixed point condition (\ref{eq:fixedpoint})
results in an operator valued recurrence relation 
\be 
(\ad H) \rho^{(p)} = -\hat{\cal D} \rho^{(p-1)}, 
\label{eq:recurrence}
\ee
for the sequence $\{ \rho^{(p)} \}$,
with the initial condition
$
\rho^{(0)}= 2^{-n} \one,
$ 
which is an infinite temperature equilibrium state.

\smallskip
\noindent
{\bf Theorem.} Solutions of (\ref{eq:recurrence}) in the first two orders read:
\begin{eqnarray}
2^n \rho^{(1)} &=& \mu (Z - Z^\dagger), \label{eq:forder} \\
2^n \rho^{(2)} &=& \frac{\mu^2}{2} (Z - Z^\dagger)^2 - \frac{\mu}{2} [Z,Z^\dagger]. \label{eq:sorder}
\end{eqnarray}
$Z$ is a non-Hermitian {\em matrix product operator}
\be 
Z = \!\!\!\!\!\sum_{(s_1,\ldots,s_n)\in\{+,-,0\}^n}\!\!\!\!\!\bra{\La} \mm{A}_{s_1}\mm{A}_{s_2}\cdots\mm{A}_{s_n}\ket{\Ra} \prod_{j=1}^n \sigma^{s_j}_j,
\label{eq:MPZ}
\ee
where $\sigma^{0}_j \equiv \one$ and $\mm{A}_0, \mm{A}_\pm$ are triple of near-diagonal matrix operators acting on an auxiliary Hilbert space
${\cal H}$ spanned by the orthonormal basis $\{ \ket{\La},\ket{\Ra},\ket{1},\ket{2},\ldots \}$:
\begin{eqnarray}
\mm{A}_0 &=& 
\ket{\La}\bra{\La} + \ket{\Ra}\bra{\Ra} + \sum_{r=1}^\infty \cos\left(r\lambda\right) \ket{r}\bra{r}, \nonumber \\
\mm{A}_+ &=& \ket{\La}\bra{1} + c \sum_{r=1}^\infty \sin\left(2\left\lfloor \frac{r\!+\!1}{2}\right\rfloor \lambda\right) \ket{r}\bra{r\!+\!1},\label{eq:explicitA}\\
 \mm{A}_- &=& \ket{1}\bra{\Ra} - c^{-1}\sum_{r=1}^\infty \sin\left(\!\left(2\left\lfloor \frac{r}{2}\right\rfloor\!+\!1\right)\lambda\right)\ket{r\!+\!1}\bra{r}, \nonumber
\end{eqnarray}
where $\lambda = \arccos \Delta \in \RaR \cup \ii\RaR$ and $\lfloor x \rfloor$ is the largest integer not larger than $x$.
Constant $c\in\CC-\{0\}$ is arbitrary, but we adopt a choice $c=1$ for $|\Delta| \le 1$ ($\lambda\in \RaR$) and $c=\ii$ for $|\Delta| > 1$ ($\lambda\in \ii\RaR$) making the matrices (\ref{eq:explicitA}) {\em always real}.

\smallskip
\noindent
{\bf Proof.} {\em First order.} From (\ref{eq:defD}) it follows that $\hat{\cal D}_\pm \one = \pm 2 (\sigma^{\rm z}_1 - \sigma^{\rm z}_n)$, and
$\hat{\cal D} \one = 2\mu (\sigma^{\rm z}_1 - \sigma^{\rm z}_n)$, hence the first order (\ref{eq:forder}) satisfies (\ref{eq:recurrence}) for $p=1$ if
\be
[H,Z] = -\sigma^{\rm z}_1 + \sigma^{\rm z}_n.
\label{eq:HZ}
\ee
It is important to observe that the ansatz (\ref{eq:MPZ}) does not contain any $\sigma^{\rm z}_j$ operator, while $[H,Z]$ can only contain terms with a single $\sigma^{\rm z}_j$.  Let us write $H=\sum_{j=1}^{n-1} h_{j,j+1}$ with $h_{j,j+1}=2\sigma^+_j \sigma^-_{j+1} + 2\sigma^-_j \sigma^+_{j+1} + \Delta \sigma^{\rm z}_j \sigma^{\rm z}_{j+1}$. 
Eq.  (\ref{eq:HZ}) implies that all the terms of $[H,Z]$ where $\sigma^{\rm z}_j$ appear in the bulk
 $1 < j < n$ should vanish. Looking locally at each triple of sites around $j$ this means
\begin{eqnarray*}
\sum_{s_1,s_2,s_3\in\{\pm,0\}}\!\!\!\!\!\!\tr\left( \sigma^{r_1}_{j-1}\sigma^{\rm z}_j \sigma^{r_3}_{j+1} [h_{j-1,j}+h_{j,j+1},\sigma^{s_1}_{j-1}\sigma^{s_2}_{j}
 \sigma^{s_3}_{j+1}]\right)\\ \times \mm{A}_{s_1} \mm{A}_{s_2} \mm{A}_{s_3}=0, \quad {\rm for\; all}\quad r_1,r_3\in\{\pm,0\}
 \end{eqnarray*} resulting in eight independent 3-point algebraic relations
 \begin{eqnarray}
&&[\mm{A}_0,\mm{A}_+\mm{A}_-] = 0,\quad \{ \mm{A}_0,\mm{A}^2_+\} = 2\Delta \mm{A}_+\mm{A}_0\mm{A}_+, \label{eq:algebra}\\
&&[\mm{A}_0,\mm{A}_-\mm{A}_+] = 0,\quad \{ \mm{A}_0,\mm{A}^2_-\} = 2\Delta \mm{A}_-\mm{A}_0\mm{A}_-,  \nonumber \\
&&2\Delta [\mm{A}_0^2,\mm{A}_+]  = [\mm{A}_-,\mm{A}_+^2],\quad 2\Delta [\mm{A}_0^2,\mm{A}_-] = [\mm{A}_+,\mm{A}^2_-], \nonumber \\
&&2\Delta\{\mm{A}^2_0,\mm{A}_+\} - 4\mm{A}_0\mm{A}_+\mm{A}_0 = \{\mm{A}_-,\mm{A}^2_+\} - 2\mm{A}_+\mm{A}_-\mm{A}_+, \nonumber \\
&&2\Delta\{\mm{A}^2_0,\mm{A}_-\} - 4\mm{A}_0\mm{A}_-\mm{A}_0 = \{\mm{A}_+,\mm{A}^2_-\} - 2\mm{A}_-\mm{A}_+\mm{A}_-.\nonumber
\end{eqnarray}
Sufficient additional conditions that the boundary terms of $[H,Z]$ containing $\sigma^{\rm z}_1,\sigma^{\rm z}_n$ result in the exact RHS of (\ref{eq:HZ}), when $Z$ is of the form (\ref{eq:MPZ}), can be expressed in terms of auxiliary boundary vectors $\bra{\La}, \ket{\Ra}$
\begin{eqnarray}
\bra{\La}\mm{A}_-&=& \bra{\La}\mm{A}_+\mm{A}_-\mm{A}_+ = \bra{\La} \mm{A}_+\mm{A}_-^2 = 0,\nonumber \\
\mm{A}_+\ket{\Ra} &=& \mm{A}_-\mm{A}_+\mm{A}_-\ket{\Ra} = \mm{A}_+^2\mm{A}_-\ket{\Ra} = 0, \label{eq:boundary}\\
\bra{\La}\mm{A}_0 &=& \bra{\La}, \quad \mm{A}_0\ket{\Ra} = \ket{\Ra}. \quad \bra{\La}\mm{A}_+\mm{A}_-\ket{\Ra} = 1. \nonumber
\end{eqnarray}
Verifying (\ref{eq:algebra}) and (\ref{eq:boundary}), which imply (\ref{eq:HZ}), for the representation (\ref{eq:explicitA}) results in trivial trigonometric identities.

{\em Second order.} To prove that (\ref{eq:sorder}) satisfies (\ref{eq:recurrence}) for $p=2$ it is sufficient to show $[H,[Z,Z^\dagger]]=(\hat{\cal D}_+ + \hat{\cal D}_-)(Z - Z^\dagger)$ (a) and 
$[H,(Z-Z^\dagger)^2] = -(\hat{\cal D}_+ - \hat{\cal D}_-)(Z - Z^\dagger)$ (b). These relations are implied by
$[Z,\sigma^{\rm z}_1-\sigma^{\rm z}_n] = (\hat{\cal D}_+ + \hat{\cal D}_-)Z$ (a') and
$\{Z,\sigma^{\rm z}_1-\sigma^{\rm z}_n\} = \frac{1}{2}(\hat{\cal D}_+ - \hat{\cal D}_-)Z$ (b'),
using the identity $[H,AB] = [H,A]B + A[H,B]$, together with (\ref{eq:HZ}), which we have just proven, Hermitian (anti)symmetrization, and the property 
$(\hat{\cal D}_\pm Z)^\dagger = \hat{\cal D}_\pm Z^\dagger$. As $\hat{\cal D}_\pm$ and (anti)commutator with $\sigma^{\rm z}_1 - \sigma^{\rm z}_n$ only act on sites $1$ and $n$, and since $Z$ on these sites can only contain terms with
 $\sigma^{+,0}_1$ and $\sigma^{-,0}_n$, respectively (see (\ref{eq:boundary})), it is sufficient to check (a') and (b') for all $Z \in\{ \one, \sigma^+_1, \sigma^-_n, 
 \sigma^+_1 \sigma^-_n\}$, which reduce to trivial algebra. QED.
\medskip 

The theorem constitutes the rigorous part of this Letter. Its essential ingredient, namely the conservation law property (\ref{eq:HZ}), shall be applied later to the long standing problem of the spin Drude weight. However, for establishing the existence of {\em a unique} NESS $\rho_\infty$ several mathematical issues still have to be addressed:
(i) It is s not clear a-priori whether recurrence (\ref{eq:recurrence}) should have a unique solution. Suppose uniqueness can be proven up to some order $\rho^{(p-1)}$.
Then, the solution of a linear equation (\ref{eq:recurrence}) is certainly non-unique up to the addition of an arbitrary linear combination of conserved quantities
$Q_k, [H,Q_k]=0$, namely $\tilde{\rho}^{(p)} =  \rho^{(p)} + \sum_k \alpha_k Q_k$. However, as has been checked explicitly by means of computer algebra for small $n \le 6$ and any $\Delta$, a unique set of coefficients $\alpha_k$ exists such that $\hat{\cal D}\tilde{\rho}^{(p)}$ is in the image of $\ad H$, and so the solution of (\ref{eq:recurrence}) for the next order $\rho^{(p+1)}$ exists. For general $n$ this statement remains a conjecture. 
(ii) For (\ref{eq:sorder}) to give the complete second order of NESS, one thus needs to check in addition that 
$\hat{\cal D}\rho^{(2)} \in \im\ad H$. This has been verified explicitly again only for $n\le 6$, and is conjectured to be true for {\em any} $n$.
(iii) Convergence properties of perturbation series (\ref{eq:pertansatz}) are unknown, although it should be useful even only as an asymptotic series.

{\em Computation of observables.} We note that for any finite $n$, due to their near-diagonality the matrices (\ref{eq:explicitA}) can be truncated to the $d=2+\lfloor n/2\rfloor$ dimensional auxiliary space ${\cal H}_d$ spanned by vectors $\{\ket{\La},\ket{\Ra},\ket{1},\ldots,\ket{\lfloor n/2\rfloor}\}$ only, reproducing exactly the same matrix product operator (\ref{eq:MPZ}). Let us now describe how to compute physical observables in NESS, 
$\ave{A} = \tr \rho_\infty A$, to second order in $\varepsilon$. For example, the spin current from site $j$ to $j+1$, 
$J_j = \ii (\sigma^+_j \sigma^-_{j+1} - \sigma^-_j \sigma^+_{j+1})$ results directly from (\ref{eq:boundary}), 
 $
 \ave{J_j} = {\textstyle\frac{1}{2}}\mu \varepsilon.
$
More interesting are expectations of local magnetization and spin-spin correlations
which are, since $Z$ does not contain any $\sigma^{\rm z}_j$, nonvanishing only in the second order in $\varepsilon$, hence we define 
$M_j := \ave{\sigma^{\rm z}_j}/\varepsilon^2$ and  $C_{j,k} := \ave{\sigma^{\rm z}_j \sigma^{\rm z}_k}/\varepsilon^2$. 
The multiple sums appearing in $M_j = 2^{-n-1} \tr \sigma^{\rm z}_j \left\{ \mu(1+\mu) Z Z^\dagger - 
\mu(1-\mu)Z^\dagger Z\right\}$ such as 
$\ave{\sigma^{\rm z}_j Z Z^\dagger} =\sum_{s_1,\ldots,s_n}\!\!\bra{\La}\mm{A}_{s_1}\!\!\cdots\!\mm{A}_{s_n}\ket{\Ra}^{\!2}\tr\sigma^{\rm z}_j\prod_{k=1}^n\!\sigma^{s_k}_k \sigma^{-s_k}_k$ 
can be evaluated using the {\em transfer matrices} (TMs), $\mm{\tilde T} = \sum_s \left(1-\frac{1}{2}|s|\right)\mm{A}_s \otimes \mm{A}_s$,
$\mm{\tilde V}=\frac{1}{2}\sum_s s \mm{A}_s \otimes \mm{A}_s$, where $\pm\equiv \pm 1$, as e.g.
$ M_j = \mu \bra{\La}\!\otimes\!\bra{\La}\mm{\tilde T}^{j-1} \mm{\tilde V} \mm{\tilde T}^{n-j}\ket{\Ra}\!\otimes\!\ket{\Ra}$,
and similarly for $C_{j,k}$ and higher correlations. However, evaluation of such TM products can be drastically simplified by observing that, since the matrices (\ref{eq:explicitA}) have only a single element in each row, the subspace of
diagonal vectors ${\cal K}$ spanned by $\ket{\nu}\otimes\ket{\nu}$, $\nu\in\{\La,\Ra,1,2,\ldots\}$ is invariant under the action of TMs
$\mm{\tilde T}{\cal K} \subseteq {\cal K}$,
$\mm{\tilde V}{\cal K} \subseteq {\cal K}$.
Using identification $\ket{\nu}\otimes\ket{\nu} \to \ket{\nu}$ and defining reduced TMs 
$\mm{T} = \mm{\tilde T}|_{\cal K}$, $\mm{V} = \mm{\tilde V}|_{\cal K}$, reading explicitly
\begin{eqnarray}
&\!\!\!\mm{T}& = \ket{\La}\bra{\La} +\ket{\Ra}\bra{\Ra} + \frac{1}{2}(\ket{\La}\bra{1}+\ket{1}\bra{\Ra}) \nonumber \\
&+& \sum_{r=1}^\infty \biggl\{ \cos^2\!\left( r\lambda\right) \ket{r}\bra{r} + \frac{c^2}{2}\sin^2\!\left(\!2\left\lfloor\!\frac{r\!+\!1}{2}\!\right\rfloor\!\lambda\right) \ket{r}\bra{r\!+\!1}
\nonumber\\
&&\qquad + \frac{c^{-2}}{2} \sin^2\!\left(\!\left(2\left\lfloor \frac{r}{2}\right\rfloor\!+\!1\right)\lambda\right)\ket{r\!+\!1}\bra{r}\biggr\},  \label{eq:T}\\
&\!\!\!\mm{V}& = \frac{\ket{\La}\bra{1}}{2}\!-\!\frac{\ket{1}\bra{\Ra}}{2}\!+\!\sum_{r=1}^\infty \biggl\{\!\frac{c^2}{2}\!\sin^2\!\left(\!2\left\lfloor\!\frac{r\!+\!1}{2}\!\right\rfloor\!\lambda\right)\!\ket{r}\bra{r\!+\!1}\nonumber\\
&&\qquad - \frac{c^{-2}}{2} \sin^2\!\left(\!\left(2\left\lfloor \frac{r}{2}\right\rfloor\!+\!1\right)\lambda\right)\ket{r\!+\!1}\bra{r}\biggr\},  \label{eq:V}
 \end{eqnarray}
we obtain efficient TM product expressions
\begin{eqnarray}
 M_j &=& \mu \bra{\La}\mm{T}^{j-1} \mm{V} \mm{T}^{n-j}\ket{\Ra}, \label{eq:TM}\\
 C_{j,k} &=& \mu^2  \bra{\La}\mm{T}^{j-1} \mm{V} \mm{T}^{k-j-1}\mm{V} \mm{T}^{n-k}\ket{\Ra}, \; j < k, \nonumber
\end{eqnarray}
etc, where, for any finite $n$, the operators (\ref{eq:T},\ref{eq:V}) can again be truncated to $d$-dimensional matrices over ${\cal H}_d$.
Note that with our aforementioned choice of constant $c$ the matrix elements of $\mm{T}$ are always positive.
\begin{figure}
          \centering	
	\includegraphics[width=0.95\columnwidth]{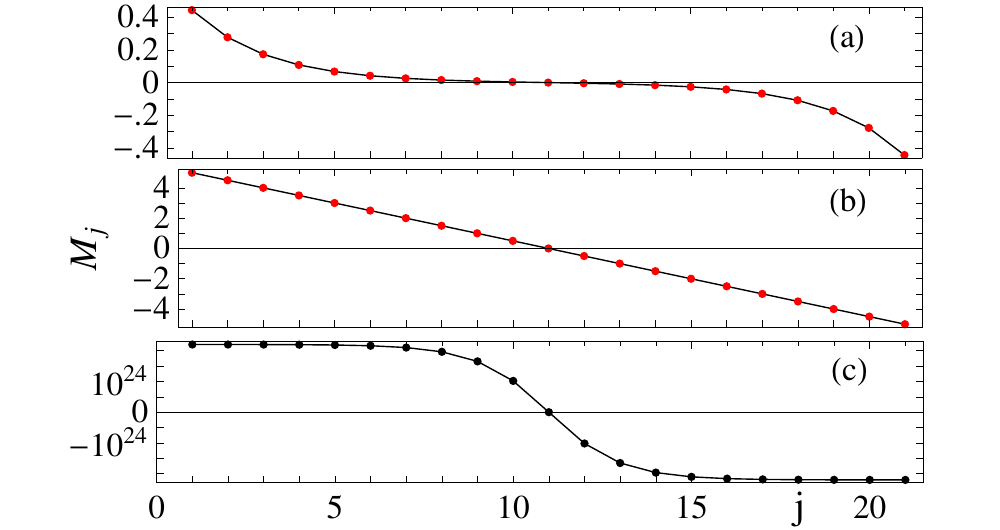}
	\vspace{-2mm}
	\caption{Scaled magnetization $M_j=\ave{\sigma^{\rm z}_j}/\varepsilon^2$ for strongly driven $\mu=1$ XXZ chain of length $n=21$ at $\Delta=0.5$ (a), $1$ (b), $1.1$ (c). Red dots mark analytical results (see text).}
	\label{fig:Prof}
\end{figure}
 
{\em Examples.} For the isotropic case $\Delta=1$, the TMs $\mm{T},\mm{V}$ have an effective rank $3$, i.e. they close on ${\cal H}_3$,
so magnetization profile is - interestingly - found explicitly to be {\em linear}, with constant long-range correlator
$M_j = \frac{\mu}{4}(n+1-2j)$, $C_{j,k} = -\frac{\mu^2}{4} ,\, j\ne k$.
Similarly, for any $\Delta$ of the form $\Delta = \cos(\pi l/ m)$, $l,m\in\ZZ$, i.e. for rational $\lambda/\pi$,
we find that ${\cal H}_{m+1}$ is invariant under $\mm{T},\mm{V}$, since the off-diagonal chains of (\ref{eq:T},\ref{eq:V}) are cut, i.e. matrix elements at position either $(m+1,m+2)$ or $(m+2,m+1)$ vanish, so $\mm{T},\mm{V}$ can be replaced by $m + 1$ dimensional matrices of $\mm{T}|_{{\cal H}_{m+1}}$, $\mm{V}|_{{\cal H}_{m+1}}$, 
independently of the system size $n$.
This allows us to find explicit results for some small $m$, say for $\Delta = 1/2 = \cos \pi/3$,
$M_j = \frac{\mu}{9}(4(5/8)^{j-1}-4(5/8)^{n-j} - (-8)^{1-n}((-5)^{j-1} - (-5)^{n-j}))$.
For $|\Delta| > 1$, however, the TMs have always infinite rank, with exponentially growing matrix elements, but (\ref{eq:TM}) allow for stable and efficient numerical computations, say of $M_j$ in only ${\cal O}(n^2)$ steps.
In Fig.~\ref{fig:Prof} we plot magnetization profiles, ranging from flatlike (for $|\Delta| < 1$) indicative of ballistic transport to kink-shaped
(for $|\Delta| > 1$) with superexponential (in $n$) growth of $M_j$ near the baths $j=1,n$ indicating the superexponential shrinking of the perturbative border
for $\varepsilon$,  and consequently also of the spin current $\ave{J_j}=\frac{1}{2}\mu\varepsilon$, since $|\ave{\sigma^{\rm z}_j}| = \varepsilon^2 |M_j| < 1$.
Note that such kink-shaped profiles $M_j$ have been observed numerically for strong driving $\mu=1$ in conjunction 
with the negative differential conductance \cite{giuliano}. One has to stress that our result, being perturbative in $\varepsilon$, cannot be used to address
the question of conductivity and spin diffusion, namely the conductivity
$\ave{J_j}/\nabla_j \ave{\sigma^{\rm z}_j} \propto \varepsilon^{-1}$ does not exist in the weak coupling limit, i.e. the limits $\varepsilon\to 0$ and $\mu\to 0$ do not commute in the {\em thermodynamic limit} (TL) $n\to\infty$.
  
{\em Mazur bound on Drude weight}. Nevertheless, our result offers very interesting physical application. Namely, as it is clear from eq.~(\ref{eq:HZ}),
the operator $Z$ commutes with the XXZ Hamiltonian, apart from the boundary terms. 
For $|\Delta| < 1$, the matrix product operator (\ref{eq:MPZ}) makes sense even for $n=\infty$, where $Z$ becomes a {\em translationally invariant pseudolocal} conservation law in the spirit of Ref.~\cite{prosen}. Since almost all matrix elements (\ref{eq:explicitA}) are smaller than $1$ for $|\Delta| < 1$, it is easy to prove that the coefficients of expansion of $Z$ in Pauli spin operator clusters decay exponentially with the length of the cluster.

\begin{figure}
         \centering	
         \vspace{-3mm}
	\includegraphics[width=0.95\columnwidth]{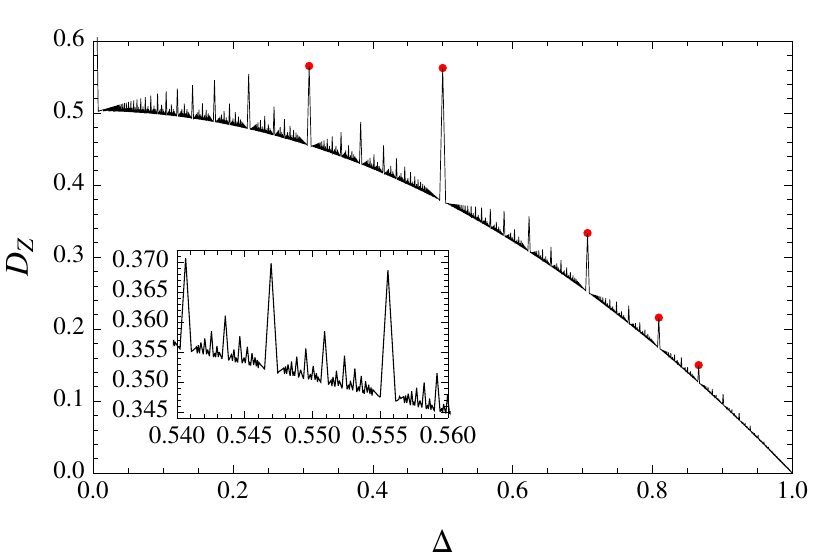}
	\vspace{-3mm}
	\caption{Mazur bound $D_Z$ on spin stiffness versus anisotropy $\Delta$ for an infinite XXZ chain, obtained by iterating 
	$(m+1)\times(m+1)$ transfer matrices at $\Delta = \cos(\pi l/m)$.
	Red dots indicate analytic results (\ref{eq:table}) for $m\le 6$. The inset magnifies $D_Z(\Delta)$ in a smaller range demonstrating its fractal nature.}
	\label{fig:D}
\end{figure}

Ref.~\cite{zotos} used the Mazur inequality \cite{mazur} to rigorously estimate the lower bound on the Drude weight $D$ at high temperature \cite{note} in terms of some conservation laws $Q_k$
\begin{equation}
D_n = \lim_{t\to\infty} \frac{\beta}{2n t}\int_0^t \dd t' \ave{J(t')J} \ge
\frac{\beta}{2n}\sum_k \frac{(J,Q_k)^2}{(Q_k,Q_k)}, \label{eq:mazur}
\end{equation}
where $J:=\sum_{j=1}^{n-1} J_j$, $J(t):=e^{\ii Ht} J e^{-\ii H t}$, $\beta$ is the inverse temperature, 
and $(A,B):=2^{-n}\tr A^\dagger B$ is Hilbert-Schmidt inner product in the operator space, which defines the norm $||A|| = \sqrt{(A,A)}$, and conserved quantities $Q_k$  are chosen to be mutually orthogonal $(Q_k,Q_l) = \delta_{k,l}||Q_k||^2$. However, for XXZ chain all the known local, or `normalizable', conservation laws \cite{mathieu} are orthogonal to the spin current $(J,Q_k) \equiv 0$, so the RHS of (\ref{eq:mazur}) vanishes.
Nevertheless, many numerical computations (e.g.\cite{fabian}) suggested that for $|\Delta| \le 1$, $D$ should be positive, although doubts have been raised recently \cite{affleck}, making the issue quite controversial. Here we propose to use the first-order-term in NESS (\ref{eq:forder}), $Q_Z:=\ii (Z - Z^\dagger)$ as Hermitian operator, which is conserved in TL, since clearly $\lim_{n\to\infty} ||[H,Q_Z]||/||Q_Z|| = 0$, and is relevant to the current, since $(J,Q_Z) = (n-1)/2$. Putting $Q_k = Q_Z$ and performing TL on both sides \cite{note2} of (\ref{eq:mazur}), we obtain 
$D = \lim_{n\to\infty} D_n \ge \frac{\beta}{4} D_Z$ where $D_Z:=\lim_{n\to\infty} (n-1)^2/[2n (Q_Z,Q_Z)]=\frac{1}{4}\lim_{n\to\infty} n/(Z,Z)$.
Computing $(Z,Z) = 2^{-n}\tr Z Z^\dagger$ by iterating TM $\mm{T}$ we get efficiently computable expression for the Mazur bound
\begin{equation}
D_Z = \frac{1}{4}\lim_{n\to\infty} \frac{ n }{\bra{\La} \mm{T}^n \ket{\Ra}}.
\label{eq:DZ}
\end{equation}
For $|\Delta|>1$, we derive $D_Z=0$ from the fact that the matrix elements in the tridiagonal band of $\mm{T}$ (\ref{eq:T}) are all positive and exponentially growing, 
which allows for numerically observed spin diffusion at infinite temperature \cite{pz,gemmer}.
Consider now $|\Delta| \le 1$. 
For a dense set of values of the anisotropy $\Delta = \cos(\pi l/m)$, TM $\mm{T}$ can be, 
as discussed earlier, replaced by a finite $m+1$ dimensional matrix $\mm{T}'=\mm{T}|_{{\cal H}_{m+1}}$, and $D_Z$ (\ref{eq:DZ}) can be evaluated {\em exactly} by means of Jordan decomposition of $\mm{T}'$. 
For example, for $\Delta=1/2=\cos\pi/3$, we find
$
\mm{T}' = \begin{pmatrix}
1 & 0 & 1/2 & 0 \cr
0 & 1 & 0 & 0 \cr
0 & 1/2 & 1/4 & 3/8 \cr
0 & 0 & 3/8 & 1/4
\end{pmatrix} = 
\mm{U} \begin{pmatrix}  
1 & 1 &   &  \cr
   & 1 &   &  \cr
   &    &  5/8 & \cr
   &   &  &\!\!\!-1/8 \end{pmatrix}
\mm{U}^{-1}.
$
Iterating $\mm{T}'$, for large $n$ the off-diagonal element of the Jordan block starts dominating, hence 
$\bra{\La}{\mm{T}'}^n\ket{\Ra} \to n \mm{U}_{1,1} (\mm{U}^{-1})_{2,2} = 4n/9$, so $D_Z(1/2) = 9/16$.
Similarly, we calculated for all $m\le 6$ (note $D_Z(\Delta)=D_Z(-\Delta)$):
\be
\begin{array}{rrrr}
\lambda & \quad D_Z(\cos \lambda) &\quad\quad \lambda &\qquad D_Z(\cos \lambda) \cr \hline
\pi/2 & 1        &  \pi/5  & 5(5-\sqrt{5})/64 \cr 
\pi/3 & 9/16  & 2\pi/5 & 5(5+\sqrt{5})/64 \cr
\pi/4 & 1/3    & \pi/6 & 3/20 
\end{array}
\label{eq:table}
\ee
For $\Delta=1$, we have $\bra{\La}{\mm{T}}^n\ket{\Ra} \propto n^2$, so $D_Z|_n \propto n^{-1} \to 0$.
For all other values of $m\le 300$ we estimated (\ref{eq:DZ}) numerically and plotted the graph of $D_Z(\Delta)$ in Fig.~\ref{fig:D} which clearly exhibits
a fractal structure. This seems to be the first appearance of fractal behavior in an infinite strongly correlated system and calls for further analysis.

Our numbers (\ref{eq:table}) have also been compared to the Bethe ansatz result \cite{zotos2} $D_{\rm BA}= 1 - \sin(2\lambda)/(2\lambda)$,
 for $\lambda=\pi/m$, which overshoots the values of the table (\ref{eq:table}) for $4\%, 9\%, 13\%$ and $15\%$ for $m=3,4,5$ and $6$, respectively. However, since the assumptions used in Ref.~\cite{zotos2} have been debated, it seems interesting to speculate whether our bound $D_Z$ may be saturating.
 
{\em Conclusion.} We developed an original method to construct explicit matrix product ansatz for the weak coupling limit of NESS in the XXZ spin chain which is driven out of equilibrium by Lindblad baths attached at the ends of the chain.
As a byproduct of this result we discovered a  pseudolocal conserved quantity which -- unlike previously known local conservation laws -- is not orthogonal to the spin current in the zero magnetization sector. Employing this conserved quantity enabled us to rigorously estimate the spin Drude weight in the metallic regime $|\Delta| < 1$ and prove the ballistic high-temperature transport there. Since the XXZ model can be mapped to a 
$t$-$V$ model of interacting fermions, our result should be equally relevant for understanding the charge transport.

We acknowledge fruitful comments by I. Affleck, T. Barthel,  J. Eisert,  E. Ilievski, M. Horvat,  P. Prelov\v sek, K. Saito, M. \v Znidari\v c
and grants P1-0044, J1-2208 of ARRS (Slovenia), as well as the Bessel prize of  Alexander von Humboldt foundation.


\begin{thebibliography}{10}
\bibitem{note} $D$ is defined in terms of the real part of optical conductivity $\sigma'(\omega) = 2\pi D\delta(\omega) + \sigma_{\rm reg}(\omega)$ and
its strict positivity in TL signals the ballistic transport.
\bibitem{note2}  In fact, the argument becomes rigorous if one first performs TL $n\to\infty$ and then the limit $t\to\infty$.
\end{thebibliography}
\end{document}